\documentclass[aip,unsortedaddress,preprint]{revtex4-1}
\draft 
\usepackage{epsf,epsfig}
\usepackage{setspace}

\begin{document}


\title{Identification of annealing temperature for high-$\kappa$-based gate oxides using differential scanning calorimetry} 




\author{Debaleen Biswas}
\affiliation{Saha Institute of Nuclear Physics, 1/AF Bidhannagar, Kolkata 700 064, India}

\author{Anil Kumar Sinha}
\affiliation{ISU, Raja Ramanna Centre for Advanced Technology, Indore 452 013, India}

\author{Supratic Chakraborty}
\affiliation{Saha Institute of Nuclear Physics, 1/AF Bidhannagar, Kolkata 700 064, India}
\email[]{e-mail: supratic.chakraborty@saha.ac.in}

\date{\today}

\begin{abstract}

\setstretch{2.0}

This article identifies the process of crystallization of thin high-$\kappa$ dielectric films and an optimal range of annealing temperature in the field of high-$\kappa$ dielectric-based metal-oxide-semiconductor (MOS) technology for its improved electrical performances. Differential Scanning Calorimetry (DSC) technique is employed to understand the thermal behaviour of thin high-$\kappa$ dielectric films of HfO$_2$, deposited by rf sputtering, on Si. The exothermic trends of the DSC signal and Grazing Incidence X-ray diffraction (GIXRD) data indicate an amorphous to crystalline transition in the high-$\kappa$ film at higher temperature. The enthalpy-temperature variation shows a glass temperature (T$_g$) at $\sim$ 590 $^o$C beyond which an amorphous to m-HfO$_2$ crystalline transition takes place. Further, the Hf-Silicate formation, observed in DSC measurement and corroborated by Fourier transformed Infrared Spectroscopy (FT-IR) studies, indicates that the process of formation of Hf-Silicate begins at $\sim$ 717 $^oC$. High-frequency (HF) capacitance-voltage $(C-V)$ and current density - voltage $(J-V)$ characteristics establish that the crystallization of the film is not the root cause of degradation of the electrical properties of the high-$\kappa$-based MOS devices, rather the device degrades due to formation of interfacial Hf-Silicate.
\end{abstract}


\maketitle 

\setstretch{2.0}

\section{Introduction}

High-$\kappa$-based dielectric materials have been studied for its use as gate dielectric, an alternative to SiO$_2$, to cope with the continuous miniaturization of metal-oxide-semiconductor (MOS) transistor structures.  The control of leakage current through the gate dielectric material has been a challenge in the field of high-$\kappa$ dielectric-based MOS technology and depends mainly on the crystallization of the high-$\kappa$ film and formation of an interfacial silicate layer.\cite{Wilk1,Wilk2,Lee,Choi,Wong,Hei} The present problems for betterment of the  high-$\kappa$-based gate dielectric devices are three folds - (i) to know the crystallization process of dielectric films; (ii) whether crystallization or Hf-Silicate formation or both are responsible for degradation of the electrical properties of the MOS device; (iii) identification of the range of annealing temperature above which the performance of the device degrades. Amorphous films, deposited by evaporation or sputtering show a glass-transition temperature.\cite{good} Availability of the above information on the cause of device degradation has a profound importance in semiconductor industry. There are conflicting views on the role of crystallization on the leakage current related properties of the device.\cite{Junga} There are very few reports that have identified the main factors responsible for degradation of the high-$\kappa$-based devices. This article identifies the process of crystallization in high-$\kappa$ dielectric films, deposited on Si, and the cause of degradation of the electrical properties of the MOS device. Among the various high-$\kappa$ gate dielectric materials, HfO$_2$ has been considered here. HfO$_2$ films are grown by various growth/deposition techniques namely, atomic layer deposition, molecular beam epitaxy, pulsed laser deposition, electron beam evaporation and sputtering techniques (dc and rf) etc.\cite{Gusev,Ho,Yan,Ikeda,Cherkaoui,Toledano,Aguirre} Further, the role of oxygen/argon ratio on the swelling of interfcial layer (IL) upon annealing in a HfO$_2$/Si-based system has been recently reported.\cite{biswas} In this article, the thermal behaviour of HfO$_2$ film on Si, deposited by rf magnetron sputtering, has been studied by DSC and accordingly, annealing temperatures were set. All the samples were studied by GIXRD and FT-IR techniques and electrical measurements were also carried out to justify the results obtained by the DSC. The results are presented and discussed here.

\section{Experiment}

$n$-type Si (100) wafer of resistivity 0.1 to 0.5 Ohm-cm, cut into several small pieces, of area of $\sim$ 1 cm$^2$ were cleaned by standard RCA method and finally a 1$\%$ HF-dip for 60 sec was performed to remove stray oxides. HfO$_2$ films were deposited on the cleaned substrates using reactive rf-sputtering system for 7.5 min with rf power of 50 Watt maintaining a deposition pressure of 5.0 mTorr, substrate temperature of 25 $^o$C and substrates rotation at 30 rpm. The Accurion-made Nanofilm EP3 ellipsometer was employed to measure the thickness of the films and the measured thickness was $\sim$ 5 nm. Further, the reflectivity study showed that the total film thicknesses for the samples upon annealing at 750 $^o$C and 550 $^o$C were respectively as 4.8$\pm$0.4 nm and 4.9$\pm$0.5 nm  which was more or less same as that of the as-deposited sample (5.0$\pm$0.6 nm).\cite{biswas} The DSC measurement was performed on one of the samples using Netzsch STA 449 C TG-DSC system. The ramp up rate was set at 10 K/min.\cite{Mazurin} The temperature was then varied from 25 $^o$C to 1150 $^o$C in N$_2$ environment. The samples were cooled at a ramp down rate of 10 K/min during the cooling cycle of DSC scan. The experiment was repeated for three times to ensure the reproducibility of the glass temperature. The following experiments were then carried out to corroborate the DSC observations. Samples were then annealed at temperatures 250 $^o$C, 550 $^o$C, 580 $^o$C, 600 $^o$C, 670 $^o$C and 750 $^o$C, set on the basis of DSC data, and were respectively denoted as A250, A550, A580, A600, A670 and A750 with as-grown sample as A. All the samples were annealed in N$_2$ ambient at the respective temperatures using Jetfirst 100 (Jipelec) rapid thermal processing (RTP) system. The GIXRD measurement on all the samples was performed at angle dispersive X-ray diffraction(ADXRD) beam line (BL-12) of Indus - 2 synchrotron source with wavelength of 0.8856 \AA using a six circle diffractometer (Huber 5020) with scintillation point detector. Perkin-Elmer-made instrument was utilized to get the FT-IR data for all the samples. Aluminium metal was then deposited on all the samples and circular electrodes of diameter of 100 $\mu m$ were patterned with a suitable mask and UV-photo-lithography using NXQ 8000 mask aligner. No post-metal annealing was performed. The HF $C-V$ measurements were then carried out at 1 MHz with E4980A LCR meter on the samples. $J-V$ measurements were also carried out on them using Keitheley 4200-SCS semiconductor characterization system. All the electrical measurements were carried out in electrically shielded and light-tight conditions.

\section{Results and Discussions}

\begin{figure}
\centerline{\includegraphics[ width= 8.5 cm ]{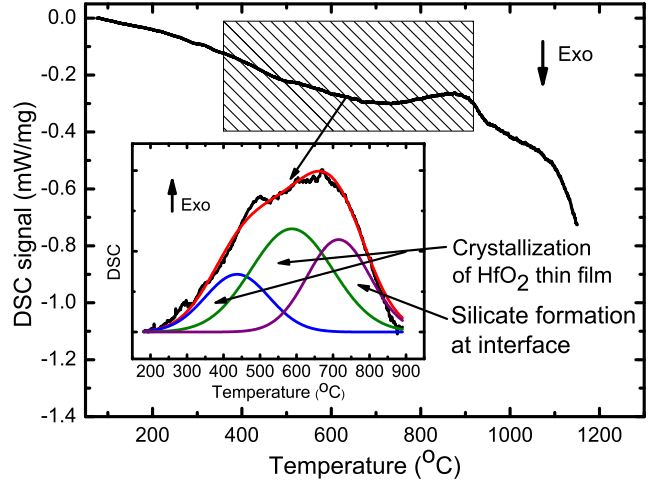}}
\caption{(Color Online) DSC signal with temperature.}
 \label{dsc1}
\end{figure}

The DSC measurement was carried out from room temperature to 1150 $^o$C and the variation of DSC signal, shown in Fig. \ref{dsc1}, illustrates a convoluted exothermic peak from $\sim$ 200 $^o$C to $\sim$ 800 $^o$C. The broad peak is fitted with three Gaussian distributions with peaks at 438 $^o$C, 587 $^o$C and 717 $^o$C, from where it is evident that two processes are involved and a two-step crystallization has been taken place.\cite{two} The first crystallization, reflected in the DSC data at lower temperature, was not observed in the x-ray study probably due to smaller fraction and size of the crystalline plane.\cite{bbb} The areas under the fitted curves i.e., the change in enthalpy during the processes, are estimated as $\sim$ 32.6 kJoule/mol and $\sim$ 20.1 kJoule/mol. The enthalpy change in amorphous to monoclinic phase transformation in the HfO$_2$ film and Hf-Silicate formation at the HfO$_2$/Si interface are reported as 32.6 $\pm$ 2 kJoule/mol and 22.3 $\pm$ 4.7 kJoule/mol, respectively.\cite{energy1,energy2}

\begin{figure}
\centerline{\includegraphics[ width= 8.5 cm ]{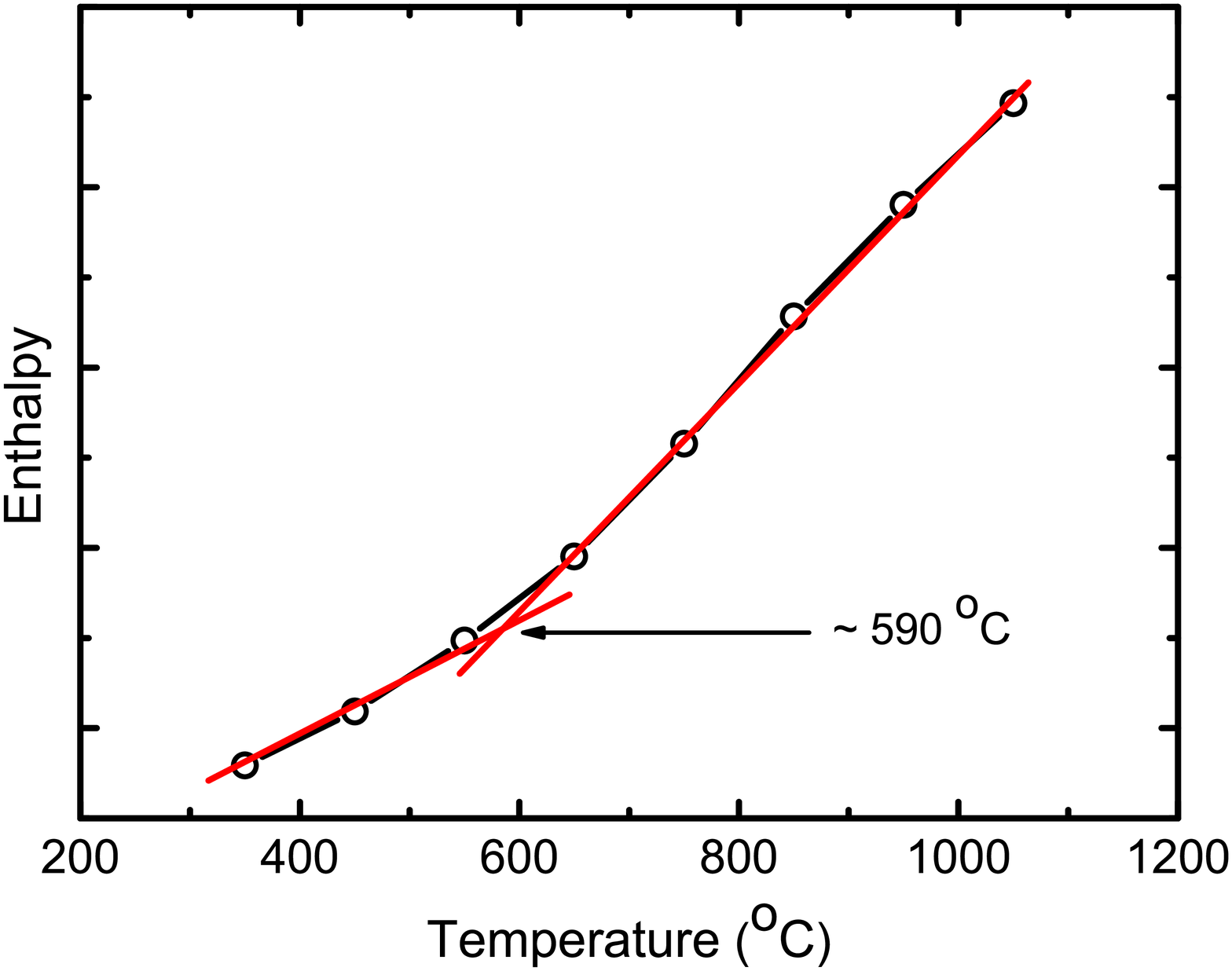}}
\caption{(Color Online) Temperature dependence of enthalpy during cooling of  HfO$_2$/Si system.}
  \label{dsc2}
\end{figure}

The variation of enthalpy with temperature is shown in Fig. \ref{dsc2}. Annealing of HfO$_2$ film changed its phase from amorphous to crystalline above a certain temperature, the glass temperature (T$_g$), which is at $\sim$ 590 $^o$C in this case. The permanent and stable crystallization phase is achieved by the HfO$_2$ film through a structural change to a long-range order from its short-range order under large mechanical stress.\cite{bbb,ccc} Significant swelling of interfacial SiO$_2$ layer (IL) and the densification of HfO$_2$ films upon annealing has been observed.\cite{biswas,rlin} As a result, a mechanical stress is developed between the IL and HfO$_2$ film due to volume expansion of SiO$_2$ IL which is obstructed by the denser top HfO$_2$ film.\cite{biswas} Since the magnitude of SiO$_2$ swelling and densification of the HfO$_2$ film strongly depend upon the annealing temperature, the magnitude of developed mechanical stress is also a function of the annealing temperature. Therefore, a critical magnitude of stress is reached at a particular annealing temperature above which a permanent crystallization of amorphous HfO$_2$ film takes place which is $\sim$ 590 $^o$C in this case. The above observation is also in good agreement with the DSC data.

\begin{figure}
\centerline{\includegraphics[ width= 8.5 cm ]{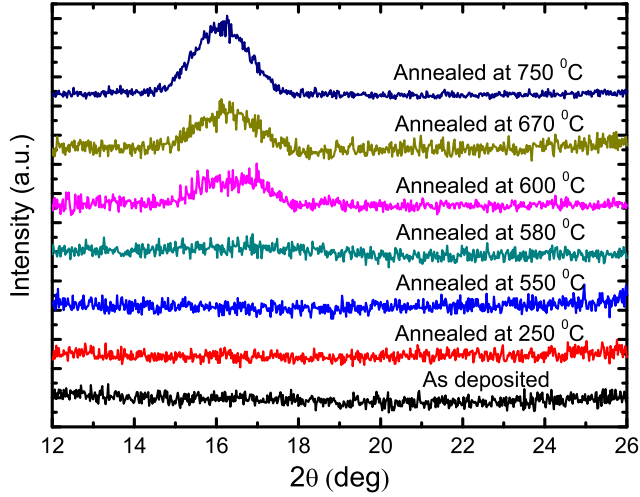}}
\caption{(Color Online) GIXRD measurements on HfO$_2$ films annealed at different temperatures.}
  \label{xrd}
\end{figure}

The GIXRD measurements on the HfO$_2$/Si system, annealed at the above temperatures, are shown in Fig. \ref{xrd}. The m-HfO$_2$ crystal plane is observed for A600 and A670 samples.\cite{pcpdf} An intense peak for the same plane is further observed for A750. Broad diffraction patterns are observed in these cases. Such broadening may be due to larger non-uniform strain, originated from thermal annealing, at the interface and reduced crystallite size.\cite{size} Interestingly, although the exothermic trend is there in the DSC data, no crystalline peaks are observed for A250, A550 and A580 samples indicating smaller fraction of crystalline plane during first step of crystallization.\cite{two}

\begin{figure}
\centerline{\includegraphics[width = 8.5 cm]{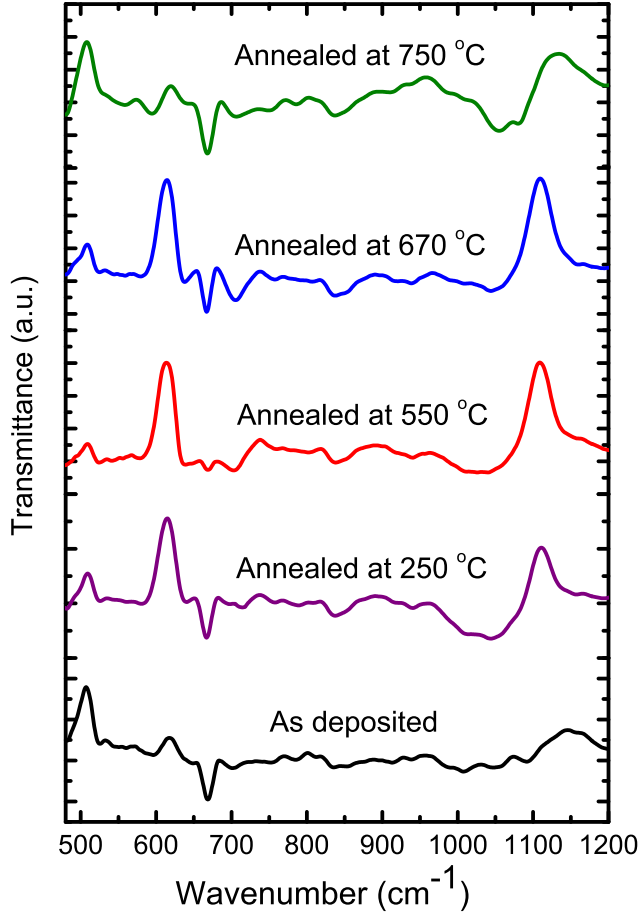}}
\caption{(Color Online) FT-IR measurements on HfO$_2$ films annealed at different temperatures.}
  \label{ftir}
\end{figure}

Fig. \ref{ftir} shows the FT-IR data for all the samples. The FT-IR data for the samples, annealed between the temperatures 600 $^oC$ and 700 $^oC$, show more or less similar and overlapped characteristics and hence, the FT-IR data of a representative sample A670 for this temperature range are presented here for discussion.  The analysis of the FT-IR data of as-deposited and annealed samples indicate the formation of Hf-O bonds and an interfacial layer.  A strong peak of Hf-O is observed at 613 cm$^{-1}$ for A250 to A670 samples.\cite{Chen} The presence of Hf-Silicate (Hf-O-Si) compound is observed for A750 sample at $\sim$ 815 cm$^{-1}$, 896 cm$^{-1}$ and 961 cm$^{-1}$. A broad peak, noticed at ~1115 cm$^{-1}$, is also due to the spectrum of silicate unit consisting of $\nu_{as}$ of longitudinal optical (LO) and transverse optical (TO) vibrations of Si-O-Si and stretching of Hf-O-Si bonds at higher annealing temperature.\cite{Toledano,Lucovsky,Neumayer,new} So, the formation of Hf-Silicate at higher temperature is clear from the FT-IR data  and the DSC scan also corroborates that the Hf-Silicate formation takes place at $\sim$ 717 $^oC$. It is already reported that at higher annealing temperature, the bottom HfO$_2$ decomposes and O$_2$ molecule is released.\cite{r18,r19} The x-ray reflectivity study further indicates that the Hf, HfO$_2$ and released O$_2$ reacts with the swelled SiO$_2$ IL at higher temperatures under stressed condition forming Hf-Silicate layer by sacrificing the SiO$_2$ IL.\cite{biswas,r23He,aaa}

\begin{figure}
\centerline{\includegraphics[width = 8.5 cm]{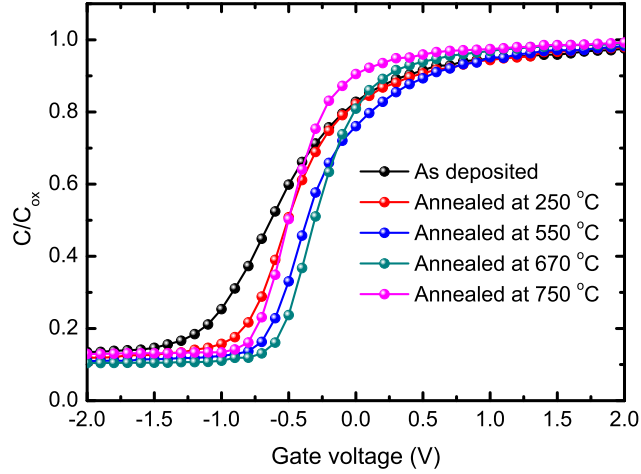}}
\caption{(Color Online) Variation of normalized capacitance with gate voltage for all devices.}
  \label{cv}
\end{figure}

\begin{figure}
\centerline{\includegraphics[width = 8.5 cm]{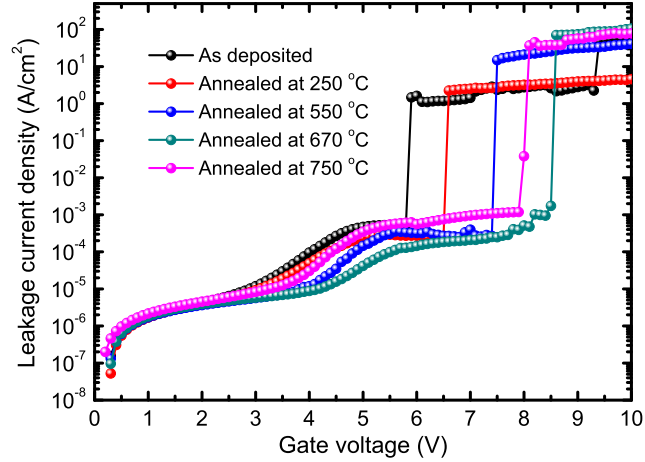}}
\caption{(Color Online) Variation of gate leakage current density with voltage for all devices.}
  \label{iv}
\end{figure}

The HF $C-V$ plots for the MOS devices are shown in Fig. \ref{cv}. The flatband voltages (V$_{fb}$) for the devices are estimated from the flatband capacitance C$_{fb}$. The HF $C-V$  and $J-V$ plots for the MOS devices, annealed between the temperatures 600 $^oC$ and 700 $^oC$, show overlapped characteristics and hence, the electrical characteristics of a  representative device A670 in this temperature range are presented here for clarity and discussion.  The flatband voltages comes out to be $-0.6$ V, $-0.5$ V, $-0.36$ V, $-0.31$ V, $-0.48$ V for the A, A250, A550, A670 and A750 devices, respectively and the corresponding estimated fixed oxide charge densities (Q$_{ox}$) are $1.1\times 10^{12}$ cm$^{-2}$, $9.8\times 10^{11}$ cm$^{-2}$, $6.2\times 10^{11}$ cm$^{-2}$, $5.2\times 10^{11}$ cm$^{-2}$ and $8.4\times 10^{11}$ cm$^{-2}$. It is observed that the Q$_{ox}$ shows a maximum for the as deposited device and improves to its minimum value for A670 device and once again deteriorates for the A750 device. The variation of leakage current density ($J$) as a function of gate voltage is depicted in Fig. \ref{iv}. It shows that the current density ramps with voltage slowly due to charge stacking in the bulk HfO$_2$.  Comparing the variations for all the samples, it is seen that upon crystallization, A670 device shows lowest leakage current. On the contrary, the leakage current density of A750 shows relatively higher value because of formation of interfacial Hf-Silicate layer along with Silicide bonds as by-product which is responsible for large defect states and oxygen vacancy that makes the leakage current properties worse.\cite{Wong,Guha} Similar trend is also observed in $J - V$ characteristics where the device breakdown takes place at higher field for the A670 device.

From the above observations, the following conclusions are made: {i) There is a glass temperature, T$_g$ above which a high-$\kappa$ dielectric film, HfO$_2$  in this case, shows a permanent crystallization. This crystallization temperature may depend on growth/deposition and annealing conditions;   ii) A critical value of mechanical stress is required for crystallization of amorphous HfO$_2$ film; iii) crystallization of high-$\kappa$ film improves the quality of the film and hence, the device performance due to an increase in its dielectric constant; iv) there is a temperature for all high-$\kappa$/Si-based MOS devices at which high-$\kappa$ silicate formation begins which is $\sim$ 717 $^oC$ in this case. This high-$\kappa$ crystallization and silicate formation temperatures may vary with growth and/or deposition conditions; v) high-$\kappa$-based silicate formation degrades the device performance due to generation of defect states that in turn increases the fixed oxide charge density. Further, the high-$\kappa$-based silicate formation reduces the dielectric strength of the MOS device causing breakdown at lower field which is evident from Fig. \ref{iv}.\cite{Hei}

\section{Conclusions}

In summary, this article is able to identify the process of crystallization of amorphous HfO$_2$ film and an optimal range of annealing temperature in the field of high-$\kappa$ dielectric-based MOS technology. A glass temperature is also detected above which the amorphous HfO$_2$ shows crystalline property. It is observed here that a critical magnitude of mechanical stress, generated upon thermal annealing, is responsible for crystallization of the HfO$_2$ film. Further, the Hf-Silicate formation at the interface/near-interface region during annealing contributes to poorer electrical properties of the devices caused by higher oxide charge and leakage current. It can be concluded that crystallization of the high-$\kappa$ film hardly contributes to the degradation of the electrical properties of the high-$\kappa$-based MOS devices. So, better electrical performance of high-$\kappa$-based device can be achieved at lower thermal budget at which the crystallization process of the high-$\kappa$ film is complete. Moreover, an optimum annealing temperature range is also suggested here but the said range will vary with dielectrics and the growth/deposition conditions.

\section*{acknowledgements}

The authors would like to thank to Prof. A. Datta, Prof. S. R. Bhattacharyya, Mr. S. Saha, Ms. M. Choudhuri of the Saha Institute of Nuclear Physics, Kolkata, India and Mr. A. Upadhyay, Mr. M. N. Singh of Raja Ramanna Centre for Advanced Technology, Indore, India for their support to this work. 


\end{document}